\magnification=\magstep1
\baselineskip=20pt
\centerline{Sub-Suns and Low Reynolds Number Flow}
\bigskip
\centerline{J. I. Katz}
\medskip
\centerline{Department of Physics and McDonnell Center for the Space
Sciences}
\centerline{Washington University, St. Louis, Mo. 63130}
\centerline{katz@wuphys.wustl.edu}
\bigskip
\centerline{Abstract}
\medskip
The phenomenon called the ``sub-Sun'' is the specular reflection of sunlight
by horizontally oriented plates of ice.  Although well-known in
meteorological optics, the hydrodynamics of the orientation is not
quantitatively understood.  I review the theory of torques on objects at low
Reynolds numbers, define coefficients $C_o$, $C_p$, and $C_\psi$ which
describe the orienting torques on discs, rods, and hexagonal prisms, and
report here the results of experiments to measure $C_o$ and $C_p$.
\vfill
\noindent
\phantom{PACS Numbers: 42.68.Ge, 47.15.Gf, 92.60.Jq, 92.60.Nv}
\eject
\centerline{I. Introduction}
\medskip
From an airplane it is sometimes possible to see a nearly specular
reflection of the Sun from a flat layer of stratus clouds, known in
meteorological optics (1--3) as a ``sub-Sun''.  This striking phenomenon is
not unusual; I have seen it twice in a few hundred flights (on most of which
the lighting or meteorological conditions were unsuitable) in the last six
years.  On the occasions on which I observed a sub-Sun (early morning in 
March over New Mexico, and late afternoon in June over Ohio) it was 
broadened by less than a degree of arc.  Where the stratus was interrupted 
by cumulus the sub-Sun disappeared, and where the stratus was partly 
transparent and a body of water appeared below it, the nearly specular 
reflection from the stratus was superimposed on the sharper specular 
reflection from the water.  
 
The well-known (1--3) explanation of the sub-Sun
is that the stratus clouds consist of small thin plates of ice maintaining a
horizontal orientation as they fall.  The angular broadening reflects a 
small dispersion in their orientations; diffraction may also contribute but 
is typically smaller.  If diffraction were dominant the sub-Sun would be 
larger in red than in blue light, appearing blue at the center but 
surrounded by a red halo, while angular dispersion of the plates leads to no
dependence on color.  I noticed no color dependence.

There are related phenomena in meteorological optics.  Light pillars (4),
like the sub-Sun, require horizontally oriented ice plates.  The
better-studied halos (5,6) may be caused by oriented or unoriented ice
crystals, either plates or columns.  The principles of the hydrodynamics of
the orientation of small falling objects are understood (7,8), but the
formidable formalism has been neither explicitly evaluated (except for
spheroids of small eccentricity and long rods) nor tested experimentally.

Under what conditions will a small, thin falling plate of ice maintain an
accurately horizontal orientation?  At high Reynolds number ${\rm Re} \gg 1$
a falling plate leaves a turbulent wake (9,10).  Its center of drag lies
close to its leading edge or surface; any steady orientation is unstable; it
tumbles, and its path is irregular because of large horizontal forces 
arising during its tumbling (ice crystals, unlike airplanes, rockets, and 
arrows, are not equipped with stabilizing tails!).  This is readily verified
by dropping a penny into a jar of water, an experiment in which ${\rm Re} 
\approx 3000$; it tumbles and usually hits the sides.  For ${\rm Re} \approx
100$ a falling disc may oscillate periodically about a horizontal
orientation as it leaves behind a regular vortex street. This may be seen by
dropping aluminum foil discs of various radii into water.  At ${\rm Re}
< 100$, however, tumbling and oscillations are strongly damped by viscosity.  
Intuitive concepts from our everyday experience with high-Re flows are still
qualitatively applicable, and show that a vertical orientation (edge-on to
the flow) is unstable; if the plate tilts the hydrodynamic force on its leading
edge acts to amplify the tilt.  However, the horizontal orientation (face-on
to the flow) is stable; if the plate tilts the wake of the leading edge
partly shields the trailing edge from the flow, reducing the drag on it;
the resulting torque restores the horizontal orientation and the disturbance
is quickly damped.
\bigskip
\centerline{II. Low Reynolds Number Flow}
\medskip
Very small plates will fall slowly, with ${\rm Re} \ll 1$.  The theory of
low-Re flow is presented by Happel and Brenner (11), who include most of the
results used in this section for flow in the limit ${\rm Re} \to 0$.  In the 
Navier-Stokes equation
$${\partial {\vec v} \over \partial t} + ({\vec v} \cdot {\vec \nabla})
{\vec v} = - {1 \over \rho}{\vec \nabla} p + {\eta \over \rho} \nabla^2
{\vec v}, \eqno(1)$$
where $\vec v$ is the fluid velocity field, $\rho$ its density, $p$ the
pressure, and $\eta$ its dynamic viscosity, the nonlinear inertial term
$({\vec v} \cdot {\vec \nabla}){\vec v} \sim v^2/\ell$, where $\ell$ is a
characteristic length, is $O({\rm Re})$ times smaller than the terms on the
right hand side, and may be neglected.  Similarly, the time derivative
term ${\partial {\vec v} \over \partial t} \sim v^2/\ell$ is of the same
order of smallness (in the absence of time-dependent external forcing), and
also may be neglected.  This leaves the creeping flow equation
$${\vec \nabla} p = \eta \nabla^2 {\vec v}, \eqno(2)$$
which is linear in velocity.  Because of the linearity of (2), linear
combinations of solutions which satisfy homogeneous boundary conditions
(such as ${\vec v} = 0$ on a solid boundary) are also solutions.

An ellipsoidal body sinking under gravity in the limit ${\rm Re} \to 0$
suffers no hydrodynamic torque, regardless of its orientation: Decompose the
fluid velocity at infinity into components along the ellipsoid's principal 
axes.  Each of these corresponds to a solution (11) of (2) which satisfies a
zero-velocity boundary condition on the body's surface, has a uniform velocity
at infinity, and (by symmetry) exerts no hydrodynamic torque about its
geometric center.  Thus the body suffers no hydrodynamic torque,
whatever its orientation, and, by the symmetry of the solutions to (2), the
center of drag is also its geometric center.  If this is also its barycenter
(as in the case of an ellipsoid of uniform density), then there is no net
torque, and all orientations are neutrally stable.  

The drag tensor ${\bf D}$ of a spheroid or triaxial ellipsoid, defined by
the relation between the applied force ${\vec F}$ and its velocity ${\vec
u}$
$${\vec F} = {\bf D} \cdot {\vec u}, \eqno(3)$$
is not isotropic.  If an external force is not directed along a principal 
axis then ${\vec u}$ will not be parallel to ${\vec F}$; in a gravitational
field an obliquely oriented ellipsoid will not sink straight down.  An
elementary calculation, using the classic results (11--13) for a circular disc
$F_{\parallel} = {32 \over 3} \eta r u_{\parallel}$ and $F_{\perp} = 16 \eta
r u_{\perp}$, where the subscripts indicate velocities and forces parallel
and perpendicular to the disc's surface, yields the result for the angle 
$\phi$ between the velocity vector and the nadir:
$$\phi = \tan^{-1} \left({\sin\theta \cos\theta \over 2 +
\sin^2\theta}\right), \eqno(4)$$
where $\theta$ is the angle between the surface normal and the vertical.
The maximum value of $\phi$ is $\tan^{-1}(1/24^{1/2}) \approx 11.5^\circ$,
which is found for $\theta = \sin^{-1}((2/5)^{1/2}) \approx 39.2^\circ$.

It is possible to generalize (in the limit ${\rm Re} \to 0$) the result of
zero torque from ellipsoids to a larger class of shapes.  The general
relation (11,14) between torque ${\vec \tau}$ and body velocity ${\vec u}$
for a nonrotating body is
$${\vec \tau} = {\bf T} \cdot {\vec u}, \eqno(5)$$
where ${\bf T}$ is a second rank pseudotensor (it changes sign under
inversion of all coordinates) because ${\vec \tau}$ is a
pseudovector.  It is not possible to construct a nonzero second rank
pseudotensor from a nonchiral geometric shape.  Hence ${\bf T} = 0$ and
there is no hydrodynamic torque on any nonchiral object in the limit ${\rm
Re} \to 0$.  It is, however, possible for the centers of drag and mass to
differ, so that the combination of hydrodynamic and gravitational forces may
lead to a net torque and a preferred orientation; a nail is an example of
such an object, which is nonetheless nonchiral.
\bigskip
\centerline{III. Not Quite Such Low Reynolds Number Flow}
\medskip
The observations of the sub-Sun clearly require that flat plates
of ice maintain a horizontal orientation, but we have seen that in the limit
${\rm Re} \to 0$ there can be no aligning torque.  The explanation is that
for any {\it finite} Re there is an aligning torque, whose magnitude may be 
estimated as first order in Re and in $\theta$:
$$\tau_\theta = C_o F_{\perp} r \theta {\rm Re}_r, \eqno(6)$$
where $C_o$ is a dimensionless coefficient of order unity applicable to 
discs and flattened oblate spheroids.  Applying $F_{\perp}$ for a 
horizontally falling disc of radius $r$ and half-thickness $h$
to a thin hexagonal plate, and using the resulting $u = \pi g r h \rho /
(8 \eta)$ and ${\rm Re}_r \equiv u r / \nu = \pi g r^2 h \rho / (8 \eta
\nu)$, where $\nu$ is the kinematic viscosity, yields
$$\tau_\theta = {\pi^2 C_o \over 4} {g^2 r^5 h^2 \rho^2 \over \eta \nu} 
\theta. \eqno(7)$$

A torque of magnitude $\tau$ acting on a disc produces a rotation rate 
(11,13,15) (about any axis!) in low Re flow ${\dot \theta} = 3 \tau_\theta /
(32 \eta r^3)$.  Using this result as an approximation for a thin hexagonal 
plate, and combining the rotational and translational flows (permitted in 
the ${\rm Re} \to 0$ limit because both satisfy the homogeneous boundary 
condition of zero relative velocity on the solid body) yields an exponential
decay time $t$ of any deviation from the horizontal orientation:
$$t = {\theta \over {\dot \theta}} = {128 \over 3 \pi^2 C_o} {\eta^2 \nu
\over g^2 r^2 h^2 \rho^2}. \eqno(8)$$
In this time the plate falls a distance $d = ut$, conveniently expressed in
terms of its radius:
$${d \over r} = {2 \over 3 C_o {\rm Re}_r}. \eqno(9)$$

A steady shear of the wind velocity produces a tilt of a falling plate
$\theta \approx v_{ij} t$, where $v_{ij}$ is an appropriate component of the
strain rate tensor of the wind shear.  In a stratus cloud the strain rate is
likely to be small or negligible if the air is stably stratified.  However, 
convective turbulence can readily disrupt the orientation of the plates, 
accounting for the observation that the specular sub-Sun disappeared 
wherever the stratus was interrupted by cumulus clouds.

The dominant influence disrupting the alignment of small plates in still air
is orientational Brownian motion (2,13--16).  The angular diffusion 
coefficient $D_o = 3 k_B T / (32 \eta r^3)$. The root-mean-square dispersion
of the angular orientation about a single horizontal axis is
$$\langle \theta^2 \rangle^{1/2} = (2 D_o t)^{1/2} = \left({8 \over
\pi^2 C_o}{k_B T \eta \nu \over g^2 r^5 h^2 \rho^2}\right)^{1/2}. \eqno(10)$$
For $r = 30\mu$, $h = 1.5\mu$ and $C_o = 1$ the dispersion $\langle \theta^2
\rangle^{1/2} \approx 0.01$ radian, roughly the largest value permitted by 
the observations.  For these dimensions $t \approx 0.2 / C_o$ sec, $u 
\approx 0.9$ cm/sec, $d \approx 0.1\ {\rm cm} \approx 30 r$, and ${\rm Re}_r 
\approx 0.02$.  Significantly smaller particles would not produce a clear 
specular reflection because of their Brownian angular dispersion, while much
larger ones would precipitate rapidly.

Ice crystals may be needle-like, and their behavior is also of interest.
The drag forces on prolate ellipsoids (11,13,15) approximate those on
needle-like hexagonal cylinders.  At low Reynolds numbers they are
$F_\parallel = 4 \pi \eta r u_\parallel / (\ln(2r/h) - {1 \over 2})$ and
$F_\perp = 8 \pi \eta r u_\perp / (\ln(2r/h) + {1 \over 2})$, where $r$ is
the longest semi-axis (half the needle's length) and $h$ is each of its 
shorter semi-axes (the needle's radius), only the leading terms in $h/r$ are
taken, and the subscripts indicate velocities and forces parallel and
perpendicular to the needle's length.  An elementary calculation of the
direction of gravitational settling yields, in analogy to equation (4),
$$\phi = \tan^{-1}\left({(1 - \zeta) \sin\theta \cos\theta \over
\sin^2\theta + \zeta \cos^2\theta}\right), \eqno(11)$$
where $\theta$ is the angle between the longest axis and the horizontal and
$\zeta \equiv (\ln(2r/h) + {1 \over 2}) / 2(\ln(2r/h) - {1 \over 2})$.
For $\zeta = {2 \over 3}$ equation (4) is recovered; plausible values of
$\zeta$ are quite close to $2 \over 3$ and lead to a very similar dependence
of $\phi$ on $\theta$.

In the limit ${\rm Re} \to 0$ needles will fall with indeterminate
orientation, by the same argument which applies to flat plates.  At small
but finite Re there will similarly be an aligning torque tending to make
their long axes horizontal
$$\tau_\theta = C_p F_\perp r \theta {\rm Re}_r, \eqno(12)$$
where the coefficient $C_p$ applies to needles and slender prolate
spheroids.  The results analogous to equations (7)--(10) are
$$\tau_\theta = {\pi C_p (\ln(2r/h) + {1 \over 2}) \over 2}{g^2 r^3 h^4 
\rho^2 \theta \over \eta \nu}, \eqno(13)$$
$$t = {16 \over 3 C_p (\ln(2r/h) - {1 \over 2})(\ln(2r/h) + {1 \over 2})}
{\eta^2 \nu \over g^2 h^4 \rho^2}, \eqno(14)$$
$${d \over r} = {\ln(2r/h) + {1 \over 2} \over 3 C_p (\ln(2r/h) - {1 \over
2}) {\rm Re}_r}, \eqno(15)$$
$$\langle \theta^2 \rangle^{1/2} = \left({4 \over \pi C_p (\ln(2r/h) +
{1 \over 2})}{k_B T \eta \nu \over g^2 r^3 h^4 \rho^2}\right)^{1/2},
\eqno(16)$$
and the angular diffusion coefficient of prolate spheroids $D_p = 3
(\ln(2r/h) - {1 \over 2}) k_B T / (8 \pi \eta r^3)$.

A needle-like ice crystal may also rotate about its long axis.  Because it
is a hexagonal prism, there are preferred values of its orientation angle
$\psi$ about this axis.  In analogy to the argument for thin plates, the
preferred orientation is that in which two prism faces (top and bottom) are
horizontal if $\theta = 0$.  The previous discussion may also be applied to 
this rotation.  Because a hexagonal prism is close to a circular cylinder, 
for its rotation about its long axis I use the relation (12) for a long 
cylinder of radius $h$ and length $2r$: ${\dot \psi} = \tau_\psi / (8 \pi r
h^2 \eta)$ (rather than that for an ellipsoid).  The torque is given by
$$\tau_\psi = C_\psi F_\perp h \psi {\rm Re}_h, \eqno(17)$$
where $h$ is used in place of $r$ both in the lever arm and in the Reynolds
number Re$_h$; the coefficient $C_\psi$ is expected to be small because of
the resemblance of a hexagonal prism to a circular cylinder, for which
$C_\psi = 0$.  The results analogous to equations (7)--(10) and (13)--(16) are
$$\tau_\psi = {\pi C_\psi (\ln(2r/h) + {1 \over 2}) \over 2}{g^2 r h^6 
\rho^2 \psi \over \eta \nu}, \eqno(18)$$
$$t = {16 \over C_\psi(\ln(2r/h) + {1 \over 2})}{\eta^2 \nu \over g^2 
h^4 \rho^2}, \eqno(19)$$
$${d \over h} = {\ln(2r/h) + {1 \over 2} \over C_\psi {\rm Re}_h},
\eqno(20)$$
$$\langle \psi^2 \rangle^{1/2} = \left({4 \over \pi C_\psi (\ln(2r/h) + {1 
\over 2})}{k_B T \eta \nu \over g^2 r h^6 \rho^2}\right)^{1/2}, \eqno(21)$$
and the angular diffusion coefficient $D_\psi = k_B T / (8 \pi \eta r h^2)$.

For an ice needle with $r = 30 \mu$, $h = 3 \mu$ and $C_p = 1$, $\langle
\theta^2 \rangle^{1/2} \approx 0.04$ radian, while $\langle \psi^2
\rangle^{1/2} \approx 0.4 / C_\psi^{1/2}$ radian.  Brownian angular
dispersion is larger for needles than for plates of comparable dimensions, 
and is particularly disruptive of the orientation of needles about their 
long axes.  However, a slender needle is lighter and falls more slowly than 
a plate whose diameter equals the needle's length, so longer needles may be 
found in clouds, and these longer needles are better oriented.
\bigskip
\centerline{IV. Experiments}
\medskip
The coefficients $C_o$, $C_p$ and $C_\psi$ can be measured, either in the
laboratory or from numerical experiments with three-dimensional
hydrodynamics calculations.  I therefore undertook experimental
measurements.  These turned out to be unexpectedly difficult, in part
because the coefficients are small.  For Re in the range 1--100 the 
orienting torques on falling discs and rods are large, and they become
horizontal very quickly (in a few diameters).  However, $C_o$, $C_p$ and
$C_\psi$ are defined as coefficients in a low-Re expansion of the torque, so
that experiments must be done for ${\rm Re} \ll 1$, for which the orienting
torques are small.  This makes the experiments sensitive to small
imperfections and heterogeneities in the shape and mass distributions of the
test objects.  The requirement that the orienting torques be large enough to
observe (and be large compared to spurious torques resulting from 
imperfections) conflicts with the requirement that the first term in an 
expansion in Re be a good approximation.  In the experiments reported here 
$0.10 < {\rm Re}_r < 0.25$, with most data obtained at the upper end of that
range.

It is easy to obtain very low Re by using very viscous fluids, but the
orienting torques would be too small to observe.  For larger Re high
viscosities imply large test objects (and increased wall effects) or higher
speeds (making observations more difficult).  Obtaining the required Re with
test bodies of reasonable size requires a quite modest viscosity; I used 
soybean oil, for which $\eta = 0.6$ poise at room temperature (17).  The
size of test bodies also represents a compromise; very small objects are
difficult to make, but large ones would require an excessively large fluid
volume to avoid wall effects on the flow, which decrease linearly in the
ratio of body size to wall distance (11); I used discs with radii in the
range 0.16--0.32 cm in a full tank of dimensions 15 cm $\times$ 20 cm
$\times$ 30 cm.  The modest viscosity requires very light test bodies to
obtain Re in the desired range; the discs were punched using ordinary paper
punches from hard 1 mil (0.0025 cm) thick aluminum foil (household aluminum
foil is softer, and discs punched from it have more ragged edges).  Data
were recorded using a hand-held 35 mm camera at about 50 cm distance.

When working with small and light bodies, surface forces are important.
Attempts to use sugar-water solutions as the fluid failed because air
bubbles invariably attached themselves to the test bodies when they were
pushed through the air-fluid surface.  This problem was not solved by the
addition of surfactants, and required the use of oil, which better wets
aluminum and pencil leads.  Even so, small bubbles of air often had to be
dislodged by stirring the fluid near them after the test bodies were
immersed.

Data were obtained from sinking discs with $r = 0.32$ cm (${1 \over
4}^{\prime\prime}$ diameter), for which ${\rm Re}_r = 0.23$, and from
discs with $r = 0.24$ cm (${3 \over 16}^{\prime\prime}$ diameter), for which
${\rm Re}_r = 0.13$.  Measurements of eight discs with $r = 0.32$ cm and of
six discs with $r = 0.24$ cm led to the results
$$C_o = \cases{0.169 \pm 0.017 & ${\rm Re}_r = 0.23$ \cr 0.26 \pm 0.05 & 
${\rm Re}_r = 0.13$.\cr} \eqno(22)$$
The (geometric) mean $\theta$ during the experiments was 16$^\circ$ for the
$r = 0.32$ cm discs and 23$^\circ$ for the $r = 0.24$ cm discs; Eq. (22)
includes a finite $\theta$ correction factor $2 \theta (\cos^2\theta + {3
\over 2}\sin^2\theta)/\sin2\theta$ which corrects for the dependence of the
vertical component of $u$ on $\theta$ and also replaces $\theta$ in Eq. 6 by
${1 \over 2}\sin2\theta$, giving the correct limiting torques at $\theta =
90^\circ$ (${\vec \tau} = 0$) and at $\theta = 0$.  

The experiments with $r = 0.32$ cm also yielded three discs whose relaxation
to horizontal was strikingly slower (by about 50\%), than that of the eight
included above (whose dispersion was about 10\%).  These outliers could have
been the consequence of unobserved small attached bubbles or manufacturing
imperfections (punching leaves a region of plastic flow and tearing around a
disc's edge which is not exactly symmetric, and is sometimes grossly
asymmetric) which displace the center of gravity from the center of drag.
If these outlying data points are included the result is $C_o = 0.154 \pm
0.031$, not significantly different than the result above.  In the
experiments with $r = 0.24$ cm two discs did not relax to the horizontal
at all and were rejected entirely; the dispersion of relaxation rates among
the remainder was greater than for the larger discs, but there were no
remaining outliers.

The errors quoted are the standard deviations of the measurements
of individual discs.  If the scatter were Gaussian with zero mean, then the
uncertainty in $C_o$ would be about 0.4 of that quoted, but this assumption
is probably unduly optimistic.  The measurements at two different Reynolds
numbers appear to disagree significantly; known sources of systematic error
are believed to be much less than the quoted errors.  This may imply that
the ${\rm Re} \to 0$ limit is not adequately approximated by Reynolds
numbers of 0.2.  It would be desirable to extend the measurements to yet
lower Reynolds numbers, but smaller discs are both more sensitive to
imperfections (because of their lower Reynolds numbers and smaller aligning
torques) and harder to observe accurately, as is shown by the greater
dispersion of the results for $r = 0.24$ cm discs.

Cylinders of mechanical pencil ``lead'' with $h = 0.0183$ cm (sold as 0.3 mm 
diameter) and lengths 0.44--0.66 cm were used to determine $C_p$.  Despite
the apparent uniformity of the cylindrical shape (their ends were flattened
by mounting them in a jeweler's lathe and grinding them against a flat
grindstone) the results were rather scattered.  The origin of the scatter may
be heterogeneity in density or remaining imperfections of the end faces.
The result was 
$$C_p = 0.059 \pm 0.016 , \eqno(23)$$ 
at a mean ${\rm Re}_r = 0.16$.  As above, the quoted error is the dispersion
of the (seven) individual measurements only.  This rough result is
less than the theoretical prediction of $C_p \approx 0.12$ for the rods used
(18,19), and again may reflect either manufacturing imperfections, unobserved
attached bubbles, or a failure to reach the ${\rm Re} \to 0$ regime at
finite $\rm Re$.  $C_\psi$ is much more difficult to measure, requiring
precision hexagonal cylinders, and I did not attempt it.

I thank D. R. Nelson and H. A. Stone for discussions, A. Biondo, R. H. 
Brazzle, J. W. Epstein, T. Hardt, R. H. Nichols and D. A. Tanner for 
technical assistance and materials, and the Office of Naval Research, DARPA
and NSF AST 94-16904 for support.
\vfil
\eject
\centerline{References}
\item{1.} M. Minnaert, {\it The Nature of Light and Color in the Open Air}
(Dover, New York, 1954).
\item{2.} R. A. R. Tricker, {\it Introduction to Meteorological Optics}
(American Elsevier, New York, 1970).
\item{3.} R. Greenler, {\it Rainbows, Halos, and Glories} (Cambridge U. 
Press, New York, 1980).
\item{4.} K. Sassen, {\it J. Opt. Soc. Am.} {\bf A4}, 570 (1987).
\item{5.} A. B. Fraser, {\it J. Opt. Soc. Am.} {\bf 69}, 1112 (1979).
\item{6.} F. Pattloch and E. Trankle, {\it J. Opt. Soc. Am.} {\bf A1}, 520
(1984).
\item{7.} R. G. Cox, {\it J. Fluid Mech.} {\bf 23}, 625 (1965).
\item{8.} W. Chester, {\it Proc. Roy. Soc. Lond.} {\bf A430}, 89 (1990).
\item{9.} W. W. Willmarth, N. E. Hawk and R. L. Harvey, {\it Phys. Fluids}
{\bf 7}, 197 (1964).
\item{10.} H.~R. Pruppacher and J.~D. Klett, {\it Microphysics of Clouds
and Precipitation} (D.~Reidel, Dordrecht, Holland, 1978).
\item{11.} J. Happel and H. Brenner {\it Low Reynolds Number Hydrodynamics} 
(Prentice-Hall, Englewood Cliffs, N. J., 1965).
\item{12.} H. Lamb, {\it Hydrodynamics} (Dover, New York, 1945).
\item{13.} H. C. Berg, {\it Random Walks in Biology} (Princeton U. Press, 
Princeton, N. J., 1983).
\item{14.} J. Garcia de la Torre and V. A. Bloomfield, {\it Quart. Rev. 
Biophys.} {\bf 14}, 81 (1981).
\item{15.} F. Perrin, {\it Jour.~de Phys.~et Radium S\'er.~7} {\bf 5}, 497
(1934).
\item{16.} F. Perrin, {\it Jour.~de Phys.~et Radium S\'er.~7} {\bf 7}, 1 
(1936).
\item{17.} R. C. Weast ed., {\it Handbook of Chemistry and Physics} 48th ed.
(Chemical Rubber Co., Cleveland, 1967).
\item{18.} R. E. Khayat and R. G. Cox, {\it J. Fluid Mech.} {\bf 209}, 435
(1989).
\item{19.} R. K. Newsom and C. W. Bruce {\it Phys. Fluids} {\bf 6}, 521
(1994).
\vfil
\eject
\end
\bye